\documentclass[twocolumn,prb,superscriptaddress,preprintnumbers,amsmath,amssymb,showpacs,floatfix]{revtex4}

\usepackage{graphicx}
\usepackage{dcolumn}
\usepackage{bm}
\usepackage{color}
\usepackage{multirow}

\newcommand{\led}{$L_{2,3}$}

\newcommand{\Fe}{CaBaFe$_4$O$_7$}

\newcommand{\Fez}{Fe$^{2+}$}
\newcommand{\Fed}{Fe$^{3+}$}

\begin{document}

\title{Orbital occupation and magnetic moments of tetrahedrally coordinated iron in CaBaFe$_4$O$_7$}
\author{N.~Hollmann}
 \affiliation{II. Physikalisches Institut, Universit\"{a}t zu K\"{o}ln,
 Z\"{u}lpicher Str. 77, 50937 K\"{o}ln, Germany}
\author{Z.~Hu}
 \affiliation{II. Physikalisches Institut, Universit\"{a}t zu K\"{o}ln,
 Z\"{u}lpicher Str. 77, 50937 K\"{o}ln, Germany}
 \affiliation{Max Planck Institute for Chemical Physics of Solids,
 N\"othnitzerstr. 40, 01187 Dresden, Germany}
\author{Hua~Wu}
 \affiliation{II. Physikalisches Institut, Universit\"{a}t zu K\"{o}ln,
 Z\"{u}lpicher Str. 77, 50937 K\"{o}ln, Germany}
\author{M.~Valldor}
 \affiliation{II. Physikalisches Institut, Universit\"{a}t zu K\"{o}ln,
 Z\"{u}lpicher Str. 77, 50937 K\"{o}ln, Germany}
\author{N.~Qureshi}
 \affiliation{II. Physikalisches Institut, Universit\"{a}t zu K\"{o}ln,
 Z\"{u}lpicher Str. 77, 50937 K\"{o}ln, Germany}
\author{T.~Willers}
 \affiliation{II. Physikalisches Institut, Universit\"{a}t zu K\"{o}ln,
 Z\"{u}lpicher Str. 77, 50937 K\"{o}ln, Germany}
\author{Y.-Y.~Chin}
 \affiliation{II. Physikalisches Institut, Universit\"{a}t zu K\"{o}ln,
 Z\"{u}lpicher Str. 77, 50937 K\"{o}ln, Germany}
\author{J.~C.~Cezar}
 \affiliation{European Synchrotron Radiation Facility, Bo\^ite Postale 220, 38043 Grenoble C\'edex, France}
\author{A.~Tanaka}
 \affiliation{Department of Quantum Matter, ADSM, Hiroshima University, Higashi-Hiroshima 739-8530, Japan}
\author{N.~B.~Brookes}
 \affiliation{European Synchrotron Radiation Facility, Bo\^ite Postale 220, 38043 Grenoble C\'edex, France}
\author{L.~H.~Tjeng}
 \affiliation{II. Physikalisches Institut, Universit\"{a}t zu K\"{o}ln,
 Z\"{u}lpicher Str. 77, 50937 K\"{o}ln, Germany}
 \affiliation{Max Planck Institute for Chemical Physics of Solids,
 N\"othnitzerstr. 40, 01187 Dresden, Germany}

\date{\today}

\pacs{71.70.Ej, 71.70.Ch, 75.25.Dk, 75.30.Gw}

\begin{abstract}
CaBaFe$_4$O$_7$ is a mixed-valent transition metal oxide having
both Fe$^{2+}$ and Fe$^{3+}$ ions in tetrahedral coordination.
Here we characterize its magnetic properties by magnetization
measurements and investigate its local electronic structure using
soft x-ray absorption spectroscopy at the Fe $L_{2,3}$ edges, in
combination with multiplet cluster and spin-resolved band
structure calculations. We found that the Fe$^{2+}$ ion in the
unusual tetrahedral coordination is Jahn-Teller active with the
high-spin $e^2_\uparrow t^3_{2\uparrow}e^1_\downarrow$
configuration having a $x^2-y^2$-like electron for the minority
spin. We deduce that there is an appreciable orbital moment of
about $L_z=0.36$ caused by multiplet interactions, thereby
explaining the observed magnetic anisotropy. CaBaFe$_4$O$_7$, a
member of the \lq 114\rq\ oxide family, offers new opportunities
to explore charge, orbital and spin physics in transition metal
oxides.

\end{abstract}

\maketitle

\begin{figure}[b]
 \includegraphics[angle=0,width=8cm]{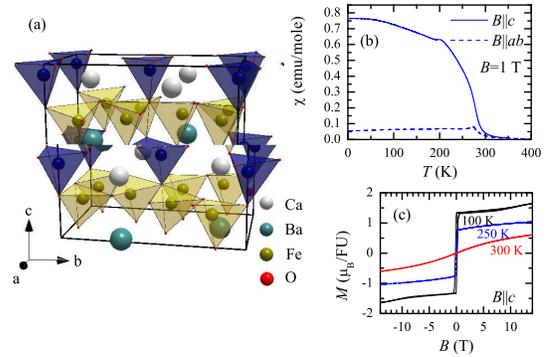}
 \caption[]{(color online) (a) A sketch of the crystal structure,
 with the FeO$_4$ tetrahedra in the kagom\'e layer (light yellow color)
 and in the trigonal layer (dark blue color). (b) Magnetic susceptibility and
 (c) magnetization of \Fe.}
\label{fig:Fechi}
\end{figure}

Transition-metal oxides are well known for their strongly
correlated electronic structure that brings orbital, spin, and
lattice degrees of freedom into close interaction.\cite{Tsuda91}
The resulting phenomena, among them high-$T_C$ and unconventional
superconductivity, colossal magneto resistance, and various kinds
of magnetic and orbital order, have been mostly studied in
compounds that are based on the perovskite
structure.\cite{Imada98} Recently, polycrystalline samples of the
new material \Fe\ were synthesized,\cite{raveau08} a compound
which belongs to the class of \emph{Swedenborgites} or \lq 114\rq\
oxides.\cite{valldor02a, valldor04a, raveau09b} It is a
mixed-valent system containing equal numbers of Fe$^{2+}$ and
Fe$^{3+}$ ions. Both ions are tetrahedrally coordinated by oxygen
atoms. This is especially very rare for an Fe$^{2+}$ ion and only
few materials containing such ions have been studied in detail
recently for their physical properties, but these are
sulfides.\cite{park99a,krimmel05a,sarkar10b} The Fe ions in \Fe\
constitute a sublattice of alternating kagom\'e and trigonal
layers as shown in Fig.~\ref{fig:Fechi} (a). Based on the shorter
average Fe-O distances,\cite{raveau08} one may expect that the
trigonal layers accommodate only Fe$^{3+}$ ions, leaving the
kagom\'e layers with a 2:1 ratio of Fe$^{2+}$ and Fe$^{3+}$ ions.
It is not known whether or not there is charge ordering in these
kagom\'e layers.

The $3d$ orbitals in a tetrahedral FeO$_4$ coordination are split
into a low $e$ and a high $t_2$ level, and the \Fez\ ion in the
high-spin state has an $e^2_\uparrow
t^3_{2\uparrow}e^1_\downarrow$ configuration. As one minority-spin
electron resides in the two-fold degenerate $e$ orbital, the
question arises whether or not this degeneracy will be lifted, and
if so, which orbital is to be occupied. The presence of such
Jahn-Teller active\cite{jahn37a,jahn38a} ions could lead to
interesting orbital physics phenomena, influencing strongly the
magnetic properties of the material.\cite{kugel82a, tokura00a} In
this work, we characterize the magnetic properties of \Fe\ by
magnetization measurements and study the orbital and spin physics
in the material using a combination of soft x-ray absorption
spectroscopy (XAS) and multiplet cluster calculations, supported
by additional spin-resolved band structure calculations.

We have succeeded to grow high quality single crystals of \Fe\
using the floating zone method. Details on the crystal synthesis
will be given in a separate paper.\cite{valldor10a} The XAS
experiments were performed at the ID08 beam line at the ESRF in
Grenoble, France. The energy resolution was $\approx 0.25$~eV at the Fe-\led\ edge, with
a degree of linear and circular polarization each higher than 99\%. The
single crystals of \Fe\ were cleaved \emph{in situ} in ultra-high
vacuum in the low 10$^{-10}$~mbar range. The Fe-\led\ XAS spectra
were taken in total electron yield (TEY) mode. Additional
measurements have been made on the O $K$ edge in the fluorescence
yield (FY) mode. The close similarity found between the TEY and FY
spectra is evidence that the TEY spectra are truly representative
for the bulk material and that the cleaved surfaces are of high
quality. Polycrystalline Fe$_2$O$_3$ was measured simultaneously
as reference. A high-field magnet with $B=5$ T was used to measure
the magnetic circular dichroism (XMCD) in the XAS spectra. The
bulk magnetic susceptibility has been measured using a vibrating
sample magnetometer in a Physical Properties Measurement System
(PPMS) of Quantum Design.

The refinement of single crystal diffraction data yields that the
crystal structure is orthorhombic ($Pbn21$, with lattice
parameters 6.3135 \AA, 11.0173 \AA, 10.3497 \AA).\cite{valldor10a}
In the magnetic susceptibility we observe magnetic ordering at
270~K similar to the data published on polycrystalline samples in
Ref.~\onlinecite{raveau08}, but in addition the single crystals
exhibit a second ordering temperature at 200~K indicated by a cusp
in the curves (Fig.~\ref{fig:Fechi}). The differences between
field- and zero-field-cooled measurements are small, indicating
the absence of magnetic degeneracy. The data also show a clear
anisotropy with the magnetic easy axis along the $c$ direction of
the crystal. The magnetic ordering may be described as \lq
ferrimagnetic\rq\ like Fe$_3$O$_4$, in the sense that the kagom\'e
layer and the trigonal layer both show canted antiferromagnetic
moments that do not compensate. The strong pinning of the moments
along the $c$ axis seems surprising, as one would expect that the
high-spin Fe ions with their half-filled $t_{2g}$ shell (\Fed\
$e^2t_2^3$ and \Fez\ $e^3t_2^3$) should not carry an orbital
moment. Yet we will show below that how the spin-obit coupling via
multiplet interactions can generate an appreciable orbital moment,
thereby explaining the observed magnetic anisotropy.

\begin{figure}[t]
 \includegraphics[angle=0,width=7.5cm]{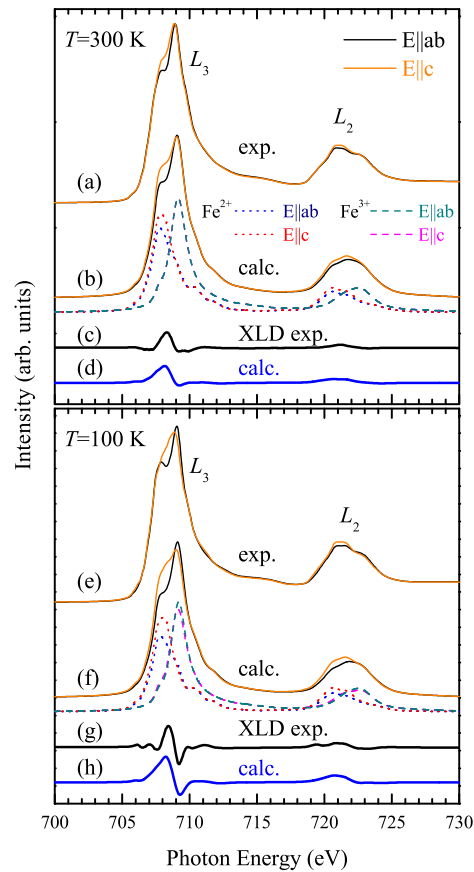}
 \caption[]{(color online) Experimental and calculated Fe \led\
  XAS spectra of \Fe\ with the $\mathbf{E}$ vector of the light
  parallel and perpendicular to the crystallographic $c$
  axis, taken at 300~K, in the paramagnetic phase (top panel) and
  at 100~K, in the magnetically ordered phase (bottom panel).}
\label{fig:FeXLD}
\end{figure}

Fig.~\ref{fig:FeXLD} shows the Fe-$L_{2,3}$ XAS spectra of \Fe\
taken with $\mathbf{E}$$\parallel$$c$ and $\mathbf{E}$$\parallel$$ab$. The top
panel displays the spectra taken at 300 K, in the paramagnetic
phase, while the bottom panel depicts the spectra at 100 K, in the
magnetically ordered phase. There is a distinct
polarization dependence, which is more pronounced for the
magnetically ordered phase. For each temperature, the difference
between the two polarizations, labeled XLD (x-ray linear
dichroism), is also plotted.

The spectra in Fig.~\ref{fig:FeXLD} are dominated by the Fe $2p$ core-hole
spin-orbit coupling (SOC) which splits the spectrum roughly in two parts, namely
the $L_{3}$ ($h\nu \approx$ 705-713~eV) and $L_{2}$ ($h\nu
\approx$ 718-725~eV) \emph{white line} regions. The line shape strongly
depends on the multiplet structure given by the Fe 3$d$-3$d$ and
2$p$-3$d$ Coulomb and exchange interactions, as well as by the
local crystal fields and the hybridization with the O 2$p$
ligands. Unique to soft XAS is that the dipole selection rules are
very sensitive in determining which of the 2$p^{5}$3$d^{n+1}$
final states can be reached and with what intensity, starting from
a particular 2$p^{6}$3$d^{n}$ initial state ($n=5$ for Fe$^{3+}$
and $n=6$ for Fe$^{2+}$).\cite{degroot94a,tanaka94a,thole97a} This
makes the technique extremely sensitive to the symmetry of the
initial state, e.g. the crystal field states of the ions.

The $L_{2,3}$ XAS process is furthermore charge sensitive, and the
spectra of higher valencies appear at higher photon energies by
about 1 eV per difference in electron occupation.\cite{Chen90,Mitra2003} We may
therefore expect tentatively that the low-energy shoulder of the
$L_3$ white line at 708~eV is due to predominantly the \Fez, while
the higher energy peak at 709~eV comes from the \Fed\ contribution.
These tentative assignments are in fact consistent with the XLD
observations. It is known that XLD can have its origin in an
anisotropy of the orbital occupation and/or in an anisotropy of
the spin orientation of the ions under
study.\cite{thole97a,Sziszar05} For a Fe$^{3+}$ ion,
however, the spherical $^{6}A_{1}$ high-spin $3d^{5}$ configuration can only
produce a significant XLD if macroscopically there is a net spin axis
present.\cite{Kuiper93} Therefore, the absence of XLD for the 709 eV peak ($\sim$\Fed) at 300~K
 and its appearance at 100~K, are consistent
with an XLD signal caused by a magnetic ordering of the Fe$^{3+}$
ions in going from 300~K to 100~K.

To check the assignments made above, we performed quantitative
simulations of the XAS spectra using the
configuration-interaction cluster model that includes full
atomic-multiplet theory.\cite{degroot94a, tanaka94a, thole97a} In
these simulations, the multipole parts of Coulomb interaction and
the single-particle SOC constant were estimated
from Hartree-Fock values that were reduced to 80\%. For the
monopole parts of the Coulomb interaction $U_{dd}$ and $U_{pd}$,
and the charge-transfer energy $\Delta$, typical values were
used;\cite{tanaka94a} with the Slater-Koster
formalism\cite{slater54a} and estimations from Harrison's
description\cite{harrison} the hybridization between Fe and O was
taken into account.\cite{Fepara} The simulations have been carried
out using the XTLS 8.3 program.\cite{tanaka94a}

The result for the simulations are displayed in
Fig.~\ref{fig:FeXLD}. One can observe good general agreement
between experiment and calculation: each individual spectrum is well
reproduced, i.e., for each of the two polarizations as well as for
both temperatures 300~K and 100~K.
The XLD is also well described. The simulations confirm that the
$L_3$ white line at 708~eV is due to the \Fez, while the 709~eV
peak comes from the \Fed. We now discuss the origin of the
polarization dependence and its implications for the orbital
occupation and spin orientation.

\begin{figure}[t]
 \includegraphics[angle=0,width=7.7cm]{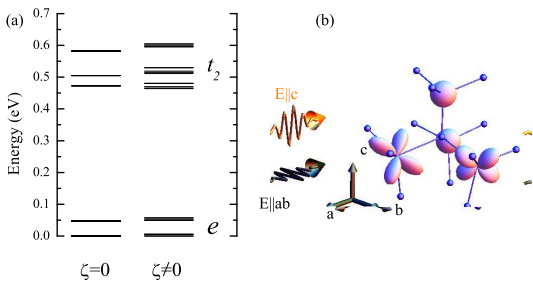}
 \caption[]{(color online) (a) Total energy level diagram for the \Fez\ ion,
  excluding (left) and including (right) the Fe $3d$ spin-orbit coupling.
  (b) An illustration of the $x^2-y^2$ orbital of the minority-spin $e$ electron of the \Fez\ ion; the high-spin half-filled
  \Fed\ ion is drawn as a sphere.}
\label{fig:FeEnergies}
\end{figure}

We start with the 300 K spectrum in which the high-spin Fe$^{3+}$
ion with its spherical $^{6}A_{1}$ 3$d^{5}$ configuration does not
contribute to the linear dichroism since there is no magnetic
ordering at this temperature. The polarization dependence is then
caused by the \Fez\ ion only and is related to its anisotropic
orbital occupation. This is revealed in Fig.~\ref{fig:FeEnergies}
(a), where we display the total energy level diagram of the ion.
Looking first at the calculation in which we artificially switch
off the $3d$ spin-orbit interaction $\zeta$, we can observe that there is
a Jahn-Teller distortion which split the $e$ and $t_{2}$
manifolds. The lowest state consists of an $e^2_\uparrow
t^3_{2\uparrow}e^1_\downarrow$ configuration with the minority-spin
electron occupying the local $x^2-y^2$ orbital, see Fig.~\ref{fig:FeEnergies}
(b). The first excited
state has a $3z^2-r^2$ like symmetry and lies 50 meV higher, i.e.
it will only be slightly occupied at 300 K. The $2p$$\rightarrow$$3d$ XAS
intensity will therefore be higher for the light polarization
vector $\mathbf{E}$ parallel to the crystallographic $c$
axis than for the perpendicular alignment as we are observing
experimentally.

We have checked this order of energy levels by carrying out band
structure calculations in the local-spin-density approximation
using the Wien2K code.\cite{wien2k} Specifically, we have set up
an impurity calculation by introducing one substitutional
Mn$^{2+}$ ion into one of the Fe$^{2+}$ sites in the kagom\'e
plane. This charge neutral substitution facilitates the
interpretation since Mn$^{2+}$ has the high-spin closed $3d^5$
shell configuration that will not lead to possibly complex
inter-orbital interactions as is in the case of \Fez. Subsequently, we have
determined the centers of gravity of the various
orbitally-resolved partial densities of states of the Mn ion. The result is that the lowest lying $3d$ orbital is of the $x^2-y^2$ type, with a $3z^2-r^2$-like orbital state lying about
50 meV higher energy. These calculations thus support our
spectroscopic analysis described above.

The orbital character of the minority-spin electron in the $e$ level is
not purely $x^2-y^2$: two effects provide mixing with other
orbitals. First, SOC mixes $t_2$ character into
the $e$ levels. This can be seen in the energy-level diagram where
multiplets that are degenerate without SOC, see
Fig.~\ref{fig:FeEnergies} (a), are split up, see
Fig.~\ref{fig:FeEnergies} (b). Second, mixing of orbital character
is also caused by the low local symmetry at the \Fez\ site as the
FeO$_4$ tetrahedra are distorted with a symmetry lower than $T_d$.
In analyzing the spectra in the magnetically ordered phase at
$T=100$~K in Fig.~\ref{fig:FeXLD} (e), one can observe that the
experimental XLD changes slightly at the position of the \Fez\ and
that an additional dichroism is created at the \Fed\ $L_{3}$ peak.
This is related to the x-ray magnetic linear dichroism (XMLD) that
is induced by the magnetic ordering.\cite{kuiper93a} In addition,
the XMLD effect not only depends on the directional axis of the
magnetic moment with respect to the polarization, but also with
respect to the local coordination.\cite{arenholz07a} The XMLD for
\Fed\ can be reproduced well by assuming the moments to be canted
by $\approx 35^\circ$ away from the $c$ axis, as seen in
Fig.~\ref{fig:FeXLD} (f) and (h). The orbital occupation is hardly
affected by the magnetic ordering.

\begin{figure}[t]
 \includegraphics[angle=0,width=7.5cm]{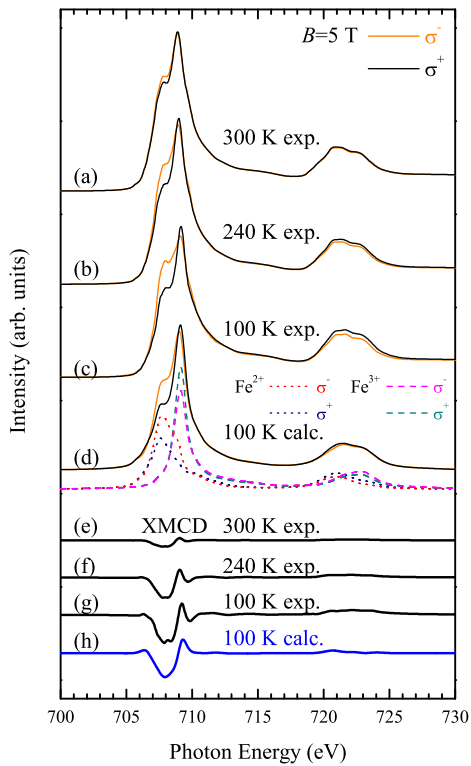}
 \caption[]{(color online) X-ray magnetic circular dichroism of \Fe.}
\label{fig:FeXMCD}
\end{figure}

We have carried out XMCD measurements of the Fe-\led\ edge, which
probe directly the local magnetism of the Fe ions.\cite{thole92,carra93,chen95} The external magnetic field was
parallel to the $c$ axis. The temperature dependence of the XMCD
signal (Fig.~\ref{fig:FeXMCD} (e)-(g)) follows clearly the
magnetic susceptibility in Fig.~\ref{fig:Fechi} (a) and (b), with a strong
increase in XMCD signal upon cooling below the first transition at
270~K, and a further increase in the second ordered phase at
100~K. Here, we can also make use of the charge sensitivity of the
XAS process: the XMCD signal at around 708~eV is mostly of \Fez\
character, the signal at 709~eV mostly of \Fed\ character. The
\lq ferrimagnetic\rq\ ordering of moments is seen in the spectra: the
sign of the signal of the two is reversed in respect to each
other. From the small XMCD signal at 709~eV we can conclude that
the two \Fed\ moments, one in the kagom\'e and one in the trigonal
plane, do not compensate and yield a small net moment. Their
net moment opposes the \Fez\ net moment, as can be seen by the
sign reversal of the XMCD signal.

Full-multiplet calculations have also been carried out to analyze
the XMCD data. We have used parameters identical to the ones used
in the simulations of the XLD spectra described above. The results
are shown in Fig.~\ref{fig:FeXMCD} (d) and (h). The agreement with the
experiment is very satisfactory, confirming that we have obtained
an accurate understanding of the local electronic structure of the
Fe ions. From these calculations, we can also extract the magnetic
properties in the form of the spin and orbital contributions to
the local moment. Along the easy axis, the expectation value of
the spin for the \Fez\ ions is $S_z=1.9$, i.e. near the
spin-only value of two. Interestingly, the \Fez\ ions each have also an
additional orbital moment of $L_z=0.36$. This is caused by
the SOC mixing the $e$ and $t_2$ orbitals, which
is not negligible since the energy difference between $e$ and
$t_2$ orbitals is rather small in tetrahedral coordination as compared to octahedral coordination. The
local distortion may cause a pronounced single ion anisotropy. As
different \Fez\ sites have their respective easy magnetic axes,
canting of the magnetic moments of \Fez\ can occur
due to these single ion anisotropies.

In conclusion, we have determined experimentally the local
electronic structure of \Fe\ using x-ray absorption spectroscopy.
Detailed information have been obtained from the linear and
magnetic circular dichroic signals in the spectra. From the
simulations using the full-multiplet cluster model, we
deduce that the \Fez\ with its high spin $e^3t_2^3$ orbital
occupation undergoes an $e$ Jahn-Teller distortion which
stabilizes a local $x^2-y^2$-like orbital state. The local
distortion of the FeO$_4$ tetrahedra and the spin-orbit coupling
mix in some $t_2$ character into this $x^2-y^2$-like state.
This induces an orbital contribution to the local magnetic
moment and causing magnetic anisotropy, which is in accordance to the bulk magnetic measurements. An arrangement of the magnetic moments was proposed.

We gratefully acknowledge the ESRF staff for providing us with
beamtime. The research in Cologne is supported by the DFG through
SFB 608. N.~H. is further supported by the Bonn-Cologne Graduate
School.

\bibliographystyle{apsrev}

\end{document}